# Internet of things and Health Care in Pandemic COVID -19: System requirments evaluation

Hasan K. NAJI
*Doctoral School of Automatic Control and Computers,*
*University POLITEHNICA of Bucharest,*
*Bucharest, Romania*
pcenghasan@gmail.com

Nicolae GOGA
*Faculty of Engineering in Foreign Languages,*
*University POLITEHNICA of Bucharest,*
*Bucharest, Romania*
*Molecular Dynamics Group*
*University of Groningen, Groningen, Netherlands*
n.goga@rug.nl

Ammar J. M. Karkar
*IT research and development centre,*
*Electronic and communication Engineering department,*
*University of Kufa,*
*Najaf, Iraq*
ammar.karkar@uokufa.edu.iq

Iuliana MARIN
*Faculty of Engineering in Foreign Languages,*
*University POLITEHNICA of Bucharest,*
*Bucharest, Romania*
marin.iulliana25@gmail.com

Haider Abdullah ALI
*Doctoral School of Automatic Control and Computers,*
*University POLITEHNICA of Bucharest,*
*Bucharest, Romania*
h_haider26@yahoo.com

*Abstract*— **Technology adoption in healthcare services has resulted in advancing care delivery services and improving the experiences of patients. This paper presents research that aims to find the important requirements for a remote monitoring system for patients with COVID-19. As this pandemic is growing more and more, there is a critical need for such systems. In this paper, the requirements and the value are determined for the proposed system, which integrates a smart bracelet that helps to signal patient vital signs. (376) participants completed the online quantitative survey. According to the study results, Most Healthcare Experts, (97.9%) stated that the automated wearable device is very useful, it plays an essential role in routine healthcare tasks (in early diagnosis, quarantine enforcement, and patient status monitoring), and it simplifies their routine healthcare activities. I addition, the main vital signs based on their expert opinion should include temperature (66% of participants) and oxygenation level (95% of participants). These findings are essential to any academic and industrial future efforts to develop these vital wearable systems. The future work will involve implementing the design based on the results of this study and use machine-learning algorithm to better detect the COVID-19 cases based on the monitoring of vital signs and symptoms.**

## I. INTRODUCTION

### A. Internet of Things in Health Care

Healthcare cost is increasing at a rapid rate, posing problems for the treatment of patients and healthcare staff because of high costs involved. Quality care provision can become difficult for patients, as well as healthcare workers [1]. Researchers identified that one solution for these challenges is using the technology to bridge the gap between practitioners and patients [2]. This strategy may also result in expanding remote and automated care. Patient and healthcare provider information exchange, communication, and assessment can be made more cost-efficient by focusing on factors like travel time, timely response, communication, and tele healthcare interventions [3]. These interventions through remote care can help the patients to reach more and access effective healthcare interventions. Moreover, it may serve more patients from broader geographic areas within the comfort of their own homes.

According to Marin et al. [1], illness prevention and patient continuum monitoring are recognised as the latest tendencies in healthcare. Thus, automated or remote care is identified among the most effective interventions that promote the objectives of patients' continuum monitoring, pandemic outbreak, and illness prevention [3]. Elderly patients are a category of population who is required to be addressed through effective solutions, such as those based on technological and automated interventions [4]. For the category who also suffers from chronic illnesses, there is a need to offer them home health solutions that are cost-efficient and use triggers, sensors, and remote monitoring systems. To efficiently implement such remote healthcare solutions, healthcare researchers are required to be addressing it through remote care and monitoring solutions [5]. However, so far, few researches have studied the possibility and the requirement of technological solution for COVID-19 specifically.

This study aims to study the value and the requirements of a remote monitoring system for patients with COVID-19. This would be through analysing the opinion. As this pandemic outbreak is growing, there is a dire need for study of such systems so that it would meets the health experts' expectations and patients' needs efficiently.

### B. Automated Healthcare Services in COVID-19

According to the report of the World Health Organization (WHO), COVID-19 has resulted in more than 2 million deaths on a global scale and over 10 million confirmed cases. Both high and low-income countries were equally affected by the pandemic. With the spread of COVID, concerns have also been identified about societal and economic effects, including global health.

The emergence of COVID related cases has been recognised as having more impact among disadvantaged and vulnerable populations [6]. Middle and low-income countries possess a fragile healthcare system. Currently, travel restrictions are still being enabled in many countries [7].

In order to improve public health workers responses to the current global COVID-19 pandemic, digital technologies are being used. Automated healthcare services are effective and provide different benefits, like contact tracing, population surveillance, evaluation of interpretations, and case identification [8]. Fast healthcare responses are empowered by millions of mobile phones, huge online datasets, increasingly cheap computing resources, machine learning techniques, and advances in language processing [9].

By using symptom-based case identification, digital systems and IoT could be a supplementary source of identification to the laboratory and clinical notification [10]. Furthermore, public health issues can be addressed by widespread access to community testing and self-testing. Existing public healthcare systems are required to be integrated with digital [11]. According to Budd et al. [10] COVID-19 cases management can be addressed efficiently through automated healthcare systems.

### C. Wearable Health Devices

Wearable Health Devices (WHDs) have been developed to help daily monitoring of vital signs for patients in ambulatory care [11]. WHDs address efficiently the gaps in clinical practices having as advantages: minimising interference and discomfort associated to normal human activities [13]. The scope of WHDs is to address and promote the concept of patient empowerment.

According to Sun, S, and, Folarin, [14], the aim of WHDs was associated with raising the interest of people, their health status, making use of new technology and enhancing patient life quality. Between several multiple science domains, WHDs create synergy by linking micro and nanotechnologies, biotechnologies, electronic engineering, materials engineering, information and communication technologies [16].

Different types of WHDs are available, particularly as smart bracelets that help to monitor patient vital signs. Clinical deterioration is efficiently addressed through monitoring different sets of valuable vital signs [17]. The five most important vital signs that are essential include heart rate, blood pressure, respiratory rate, body temperature, and blood oxygen saturation [18]. Continuous monitoring should be made for patients by analysing the previous mentioned vital signs.

E. Kańtoch and A. Kańtoch [18] state that automated vital signs data collections improve the accuracy, efficiency and completeness of clinical documentation process. Timely data collection of patient vital signs may help healthcare practitioners to develop their treatment plan more efficiently [19]. Clinicians and healthcare staff easily understand data generated from automated vital sign collection. Moreover, with the advancement in WHDs, their implementation in routine clinical practice is highly recommended [19].

### D. Aim and Objectives

In order to design a COVID-19 system that integrates a smart bracelet that helps signal a patient's vital signs, the value and functionality requirements of the wearable system should be analysed. This is achieved through conducting of the questionnaire that is designed to collect the data from health experts on system functionality requirements, which then leads to design specifics such as system architecture and choosing the types of sensors that will be used in the system and which are the most effective [25].

## II. PROPOSED COVID BRACELET VS. OTHER SOLUTIONS

There are several similar COVID-19 systems. In what follows we discuss how we are different from existing systems.

Mujawar, (2020) system investigated efficient miniaturized Nano-enabled sensors are suitable for patient's diagnostics. The smartphone assisted patients diagnostics is making personalized diagnostics possible, and managing bioinformatics and big data analytics to optimize personalized wellness. By IoT-based diagnostics is emerging for targeted disease management, we can assisted patient diagnostics for COVID-19 pandemic management.

The proposed COVID-19 bracelet is different from Tavakoli (2020) as the proposed system foreseen to have an integrated watch inside it. In addition, the proposed system consists of a temperature sensor, pulse sensor, and oximeter installed in addition to Bluetooth and battery. The fixed part, inside the house, consists of a specialised processor and GPS [21].

A device made by Weizman et al., (2020) lacked certain integrated systems, such as GSM, Bluetooth, and power, which are present in the currently proposed system.

According to Walline (2020), in China, quarantine tracker wristbands made out of waterproof paper on which was printed a serial number and a QR code. The code allows the user to access the "Stay Home Safe" mobile phone application. After installing the application, the user needs to walk around the perimeter of her/his home in order to set the boundaries of the quarantine zone.

The user monitored for 14 days during which, when they want to get out of the scanned zone, an additive warning is triggered by their mobile phone. The user needs to return to the quarantine space in maximum 15 seconds and to confirm the presence after the QR code scanned. Compared to this tracking solution, the COVID-19 bracelet proposed in the paper is not intrusive and it allows vital signs to be monitored, helping users to monitor their health.

El Majid et al. (2020) proposed a smart bracelet design, which aims to disinfect the hands of the users, as well as the objects, which are touched by them. This is done by integrating in the bracelet an intelligent electronic system that emits ultraviolet rays. Contrasted with our solution, the UV bracelet may determine the premature aging of skin, along with the appearance of wrinkles, spots, keratosis, and eye problems.

The Embrace2 bracelet from Empatica [22] was designed for persons who suffer from epilepsy and the software called Aura is responsible of tracking early respiratory infections based on physiological data regarding heart rate and skin temperature. All these features determine it to be useful for COVID-19 patients. The bracelet lacks the oximeter facility, which is present in the proposed solution.

The proposed system (The bracelet) will depend mainly on selected sensors based on the medical opinion and the questionnaire conducted in this paper and the recommendations of experts and virus specialists, in accordance with the standards and guidelines issued by the World Health Organization to monitor and follow-up of those infected with the virus over a 24-hour period.

This system will help the medical team to avoid direct contact with COVID-19 patients, reduce infection between the medical team prevent the spread of the virus and reduce the actual need for medical treatment with health institutions, especially during a pandemic. One medical member can monitor several patients in one or several rooms and record all readings in a database designed for this purpose.

The system will alert the supervising medical team when the readings are different (temperature and blood oxygen saturation). The system will need software that regulates the communication between the sensors and the digital processors.

This system can also be used on a personal level for cases of moderate and mild symptoms (that do not require hospitalization), and it is preferable to treat at home for one or more family members by direct contact with health institutions when needed [23].

### A. Participants

Healthcare practitioners are the targeted population for this quantitative study. We used purposive sampling. A number of 376 (Doctors, Pharmacists, Nurses, and Biologists) individuals belonging to different health professions participated in the research. The participants were doctors, pharmacists and lab workers. The questionnaire distributed online, being entitled *"A questionnaire to design and implement a bracelet to monitor the health status of patients with COVID-19 System requirements evaluation"*

### B. Ethics

Participants approached to fill the informed consent before recruited in the study. To keep their data confidential, participants' data protected. Furthermore, to comply with the ethics of beneficence, survey questions were ensured not to damage participants' feelings [26].

During data collection, participants could withdraw from the study at any time. The current investigation also ensured that each participant finds the questions in the survey easy to understand.

### III. RESULTS

The system intended to provide access-monitoring services for patients with mild to moderate symptoms with COVID 19 virus, or among the ones who have been in home quarantine period. The Identification of early signs and symptoms of COVID-19 that highly recommended by W.H.O in most countries to prevent the pandemic from reaching critical case. The demography of this study is wide. For instance, 42% the questionnaire participants were males and 58% were females. Thus, the majority of participants in the current investigation were females. Out of 376 participants, 66% of participants belonged from the age group 20-30, while 28.2% of participants were aged 30-40. Participants investigated for their current job and position. 90.7% were pharmacists and 7.7% were doctors. The remaining number of participants was nurses and laboratory workers.

In addition, participants were asked: what are the electronic products that are used to follow patients' vital signs. This question is to see if they are already dependent on some electronic devices.

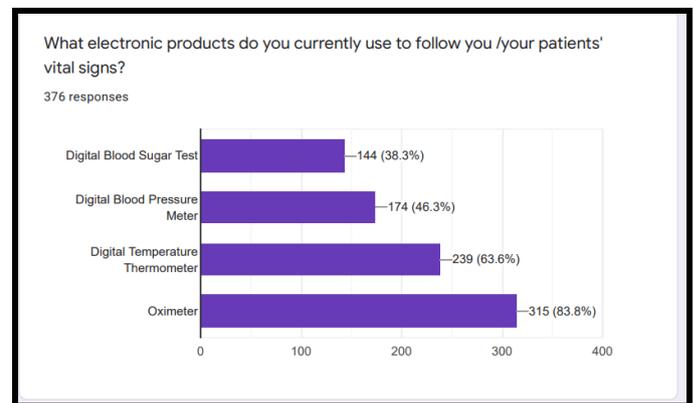

Figure 1. Electronic products usage to vital signs measurement

According to the responses (Fig. 1), 83.8% use oximeters, 63.6% use digital temperature thermometer, 46.3% use digital blood pressure meter, and 38.3% use digital blood sugar test. Oximeters used for measurement of oxygen saturation for patients with issues of breathlessness. Participants questioned about their satisfaction level with automated devices. According to the responses of participants, 66.5% choose good, 18.4% responded very good, and 13.3% responded fair. Despite the efficiencies of electronic products, the satisfaction level of the majority of participants remained at the scale of good only. Participants asked regarding the frequency of use of their mobile phones to give medical advice to patients. 53.6 % responded sometimes, 32.5% responded always, and 9.6% responded rarely. This indicates that participants do not use automated gadgets frequently.

The essential vital signs that they monitor for COVID patients also investigated. According to responses of participants (Fig. 2), 95.2% responded oxygenation level, 66.2% responded temperature, 29% responded heart rate, and 16% blood pressure.

Majority of the participants stated that oxygenation level and the body temperature are the most essential vital sign that monitored frequently followed by other vital signs including heart rate and blood pressure.

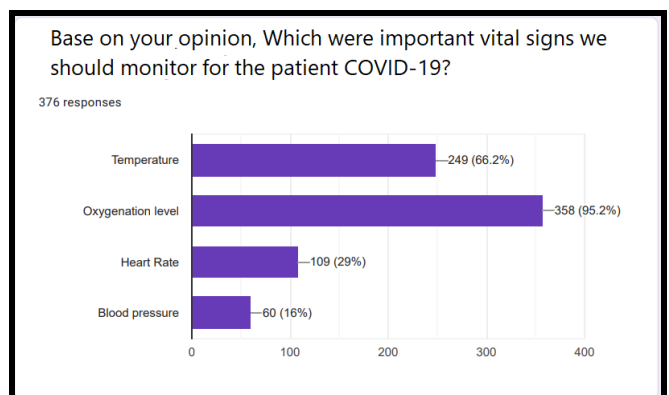

Figure 2.        Essential Vital Signs for COVID Patients

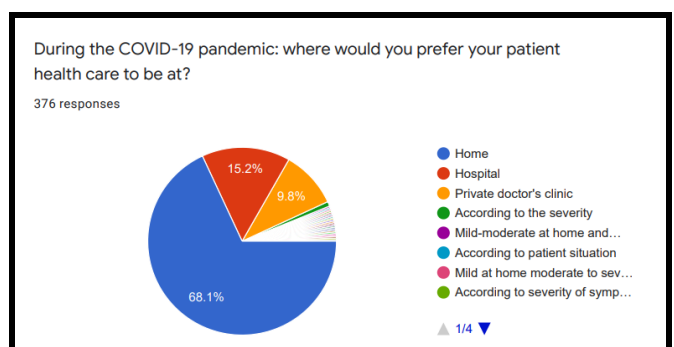

Figure 3.        Preference Regarding Care of COVID Patients

As expected, reliability and accuracy of data collected through COVID-19 device play an important role. According to participant responses, 91.2% confirmed that reliability and accuracy of data collected through COVID-19 device are important to be considered.

Participants were asked about the ease of use of COVID device. 97.9% of participants responded *important,* while remaining responded *not important*. Majority of participants consider that devices must be easy to use.

Participants were asked regarding the security of data collected through the COVID device. According to responses of participants, 55.9% responded *important*, 30.3% responded *neutral*, 12.8% responded *not important*. Participants consider that security of the COVID device data is important to be considered and maintained.

The price of the system was found to matter for 90.7% of participants, while 9.3% considered that the system price does not matter. Many participants consider that automated devices and electronic system prices significantly matter when they need to integrated into a healthcare system.

The top preferences regarding patient care (Fig. 3) were home (68.1%), hospital (15.2%), and private doctor's clinic (9.8%), followed by decisions, which depend on illness severity and situation.

Study participants were investigated whether they travelled during COVID pandemic. 80.3% participants responded positively, while 19.7% responded negatively. Moreover, they were asked if they respected or not the quarantine period after travelling.

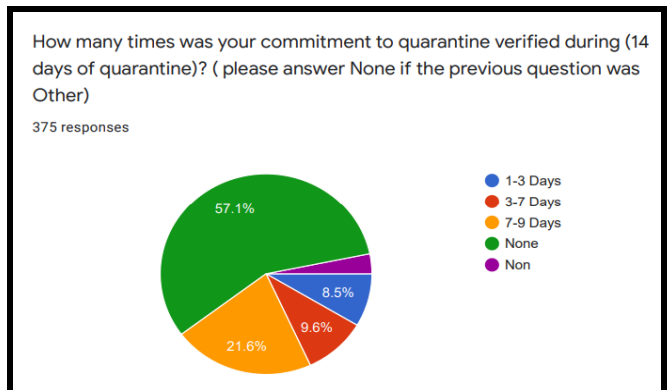

Figure 4.        Commitment to Quarantine

Based on responses, 71% answered positively, while 29% responded negatively. Findings indicate that majority of participants were aware of quarantine and COVID transmission prevention. Quarantine commitment is demanded. According to participant responses, 42.3% responded family member, and 14.1% responded family doctor. Remaining participants responded police, none, no one, others, and themselves. Findings indicate that many participants are comfortable with a family member than any other carer.

Participants were questioned regarding how many times their commitment to quarantine was verified. According to responses (Fig. 4), 57.1% stated none, 21.6% stated 7-9 days, 9.6% stated 3-7 days, and 8.5% stated 1-3 days.

## IV. DISCUSSION

The current investigation aims to identify the components and medical evaluation based on expert opinion to create medical systems to monitor patients during the pandemic, and to address the environmental and practical impacts on such systems. According to the responses of the participants, there are several types of electronic products are used in the preparation of such systems to include common electronic products used, for example, pulse oximeters, digital thermometers, digital sphygmomanometers, and digital blood sugar tests.

Electronic devices are an important part of providing healthcare services and improving patient care. The majority of automated devices used in healthcare to record and monitor vital signs. The ease of use of robotic devices is one of the most prominent features of the newly developed innovative devices. Study participants also indicated that their level of satisfaction also improved by using robotic and electronic devices to measure patients' vital signs.

In healthcare institutions, collecting and recording patient vital signs is a routine task especially in low-intensity, non-critical care hospital settings. Patient vital signs collected in general wards, which are the busiest hospital areas with large numbers of patients.

Previously, vital signs were recorded and collected at different intervals by healthcare practitioners in a manual manner, which also relied on the technical skills of health care providers. The process involves obtaining vital sign measurements; regardless of the period in which the collected data is collected and subsequently transferred to the patient's medical record manually.

The results of the questionnaire, were in the majority that temperature is an important vital sign for following up and monitoring Covid-19 patients at a rate of 66%, while the oxygen saturation sensor was 68% and it was identical to what was confirmed by the W.H.O in most of its reports of the virus. Therefore, it will be adopted by the proposed system on the basis on these products.

The heart rate and blood pressure sensors were also excluded for their low effective on the signs of disease in a compared to the temperature and oxygen sensors, and these results were in accordance with the reports issued by results of the questioner with 29 % and 16%.

The process of wireless transmitting the data and using wearable devices will give safety and flexibility to system users and health care providers, and it will give a health care indirectly lead to reducing the risk of infection spreading among medical personnel on the one hand and spreading infection among healthy people .indicate the participants opinion was 97 % important.

Wearable devices with sensors will also provide an approved complementary environment for people who prefer to receive treatment inside their homes by 66% especially those with mild and moderate symptoms, therefore, based on the search while the some experts preferred the hospitals and medical care institutions by 15% in critical case.

Based on the point of view of the experts and specialists in this field who participated in the questionnaire all this will lead to the early detection of the virus and early identification of the health problems resulting from it, which enables medical personnel to make timely decisions.

The process of using both Bluetooth and Wi-Fi as being among the approved carriers of such systems, especially by relying on the principle of privacy and security, whose importance the questionnaire confirmed is 91% to keep the health data security.

The opinion of the participants in the questionnaire regarding the level of performance, safety measures, accuracy and ease of use of the system was of clear importance through the opinions of the participants, at a rate of 98%.

The importance of the system lies in the use of certified parts and from authorized companies to give real results in addition to choosing the locations in which the sensors will be installed, the method of building the bracelet, and so on, in terms of technical and administration. The expert's confirmation that the costs of the system should be low cost, as 90 % participants emphasized the importance of the cost. The system provided that this does not lead to unreliability of the system outputs.

The table below shows the Functional requirements and Non-functional requirements of the system based on the results of this questionnaire

This obstacle we will overcome to ensure the success of the system by building a medical bracelet that matches healthcare requirements (especially with the Corvid-19 pandemic), in order to ensure that the data is transferred to the products that with ease of use and less cost and recommended from experts in that field.

This obstacle was overcome to ensure the success of the business by building a medical bracelet that matches healthcare requirements (especially with the Corvid-19 pandemic), in order to ensure that the data is transferred to the products that we do.

TABLE I.

IMPORTANT REQUIREMENTS IN WEARABLE SENSING DEVICES

| Requirements in Sensing Devices | |
|---|---|
| Functional requirements | Non-functional requirements |
| Temperature sensing | Performance |
| Oxygen level sensing | Safety |
| Blood pressure and Pulse rate sensing | Security |
| Transmitting the vital signs | Reliability |
| Early identification of health problems | Usability |
| Sending information via Bluetooth or WiFi | Flexibility |

Through the results of this investigation, we must choose the positions of the sensors carefully to ensure accurate readings, and emphasis must be placed on choosing energy sources to provide stability and continuity of operation of the system for the longest possible period, taking into account the size and effectiveness. As for the external appearance, the shape and size of the bracelet to be manufactured (3D printer) must ensure durability, safety and comfort for the patient.

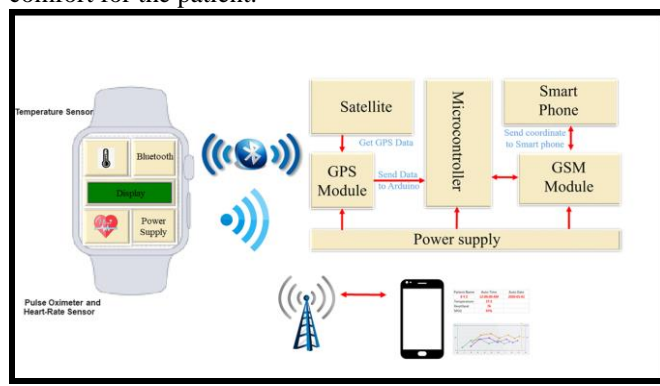

Figure 5.     Implementation Model

V. CONCLUSION

This research aimed to study the requirements to implement smart bracelet. This wearable device is critical in responses efforts to face the COVID-19 pandemic. Primary quantitative research design has been used in this investigation and an online questionnaire was distributed among a targeted population of participants. The survey investigated the essentiality and importance of COVID-19 device, a smart bracelet that helps in signalling the vital signs, for the medical team.

Research findings indicate that majority of healthcare practitioners (around 98%) find it important to use automated devices in daily healthcare to face this pandemic. Moreover, this study results clearly indicates that assessment and mentoring of vital signs (such as temperature (66%) and oxygenation level (95%)) is the major requirements of such electronic wearable devices.

In addition, a major demand of such system includes enforcing quarantine of suspected individuals at an early stage based on clinical assessment of vital signs. On the other hand, findings indicate that healthcare practitioners consider the reliability, security measures, and user data protection an important aspect that must be sustained in wearable devices and sensors.

As a result, these wearable devices are predicted to be effective in early detection and monitoring of signs and symptoms for COVID-19. This study would has determined the specifications and value of such systems and would be useful for interested industry and academic institutions alike. Future work would include implementing a smart bracelet system for the COVID-19 pandemic to monitor the healthcare status of people who suffer from chronic diseases, the elderly, and even those with special needs. Moreover, collected patient data could be analysed using data mining algorithms to extract information and automat pandemic response and/or treatment administration.